\documentclass[twocolumn,amsmath,showpacs,amsfonts,aps,prc,floatfix]{revtex4}
\usepackage{graphicx}
\usepackage{bm}
\usepackage{amssymb,amsfonts,amsmath}

\begin{document}

\title{Elliptic Flow from a Beam Energy Scan:

a signature of a phase transition to the Quark-Gluon Plasma}

\begin{abstract}
We employ a relativistic transport theory to describe the fireball expansion of the matter created in ultra-relativistic
heavy-ion collisions (uRHICs).  Developing an approach to fix locally the shear viscosity to entropy density $\eta/s$,
we study the impact of a temperature dependent $\eta/s(T)$ on the build-up of the elliptic flow, $v_2$,
a measure of the angular anisotropy in the particle production.
Beam Energy Scan from $\sqrt{s_{NN}}= \rm 62.4 GeV$ at RHIC up to 2.76 TeV at LHC has shown that the  $v_2(p_T)$ as a function of the
transverse momentum $p_T$ appears to be nearly invariant with energy. We show that such a surprising behavior
is determined by a rise and fall of $\eta/s(T)$ with a minimum at $T\sim T_c$, as one would expect if the matter
undergoes a phase transition or a cross-over. This provides an evidence for phase transition
occurring in the uRHIC's and a first constraint on the temperature dependence of $\eta/s$. In particular, a constant
$\eta/s$ at all temperatures or a too strong T-dependence would cause a breaking of the scaling of
$v_2(p_T)$ with the energy.

\end{abstract}



\author{S. Plumari$^{a,b}$, V. Greco$^{a,b}$ and L.P. Csernai$^c$\\
$^a$ \small{\it Department of Physics and Astronomy, University of Catania,}\\
\small{\it Via S. Sofia 64, I-95125 Catania (Italy)}\\
$^b$ \small{\it Laboratorio Nazionale del Sud, INFN-LNS, Via S. Sofia 63, I-95125 Catania (Italy)}\\
$^c$ \small{\it Dept. of Physics and Technology, University of Bergen (Norway)}}

\maketitle

The main motivation for the ultra-relativistic heavy-ion collisions program was to create a transient
state of a quark-gluon plasma matter (QGP) \cite{STAR_PHENIX,Adcox:2004mh}.
One of the main discoveries at RHIC was that such a matter has a very low shear viscosity
to entropy density ratio $\eta/s$ \cite{Shuryak:2003xe} close the conjectured lower bound
for a strongly interacting system in the limit of infinite coupling, $\eta/s=1/4\pi$ \cite{Kovtun:2004de}.
The first and principal observable indicating such a low viscosity is the so called elliptic flow,
a measure of the anisotropy in the angular distribution of the emitted particle defined as
$v_2= \langle cos( 2\, \varphi_p) \rangle= \langle (p_x^2-p_y^2)/(p_x^2+p_y^2) \rangle$
with $\varphi_p$ being the azimuthal angle
in the transverse plane and the average meant over the particle distribution.
When analyzed quantitatively by means of hydrodynamical simulation \cite{Kolb:2003dz}, it has been found
that the amount of $v_2(p_T)$ observed is nearly consistent with the one of a perfect fluid,
while an advanced analysis by means of viscous hydrodynamics \cite{Romatschke:2007mq,Heinz} or transport kinetic theory
\cite{Ferini:2008he,Xu:2007jv,Xu:2008av,Cassing:2009vt,Bratkovskaya:2011wp,Plumari_Bari}
both confirm that the data on $v_2$ at RHIC and LHC are consistent with an average
$4 \pi \eta/s \sim 1-3 $. More recently the possibility to measure event-by-event the angular distribution of emitted
particle has made possible the measurement of higher harmonics $v_n= \langle cos(n\, \varphi_p) \rangle$
with $n>2$ showing a fast decrease of harmonics with $n>3$ again compatible with a finite but not too large
value of $\eta/s$ \cite{Cifarelli:2012zz,Staig:2011wj}.

A low value of $\eta/s \sim 0.1$ in itself is not a direct signature of the creation of a QGP
and indeed a QGP as expected from asymptotic freedom should have an $\eta/s$
about one order of magnitude larger \cite{Arnold:2003zc}.
It is known from atomic and molecular physics that a minimum in $\eta/s$ is expected close to
the transition temperature as emphasized in the context of QGP in Refs. \cite{Csernai:2006zz,Lacey:2006bc}.
Such a minimum can be relatively smooth if one considers a system above the critical point
or more pronounced at and below the critical point with the possibility to have discontinuity \cite{Csernai:2006zz,Lacey:2006bc,
Lacey:2007na}.
For this reason not an average value for $\eta/s$, but rather a phenomenological
estimate of its temperature dependence, is desired to find a confirmation that the matter
created  undergoes a phase transition.

\begin{figure}[h]
\begin{center}
\includegraphics[width=8.1 cm]{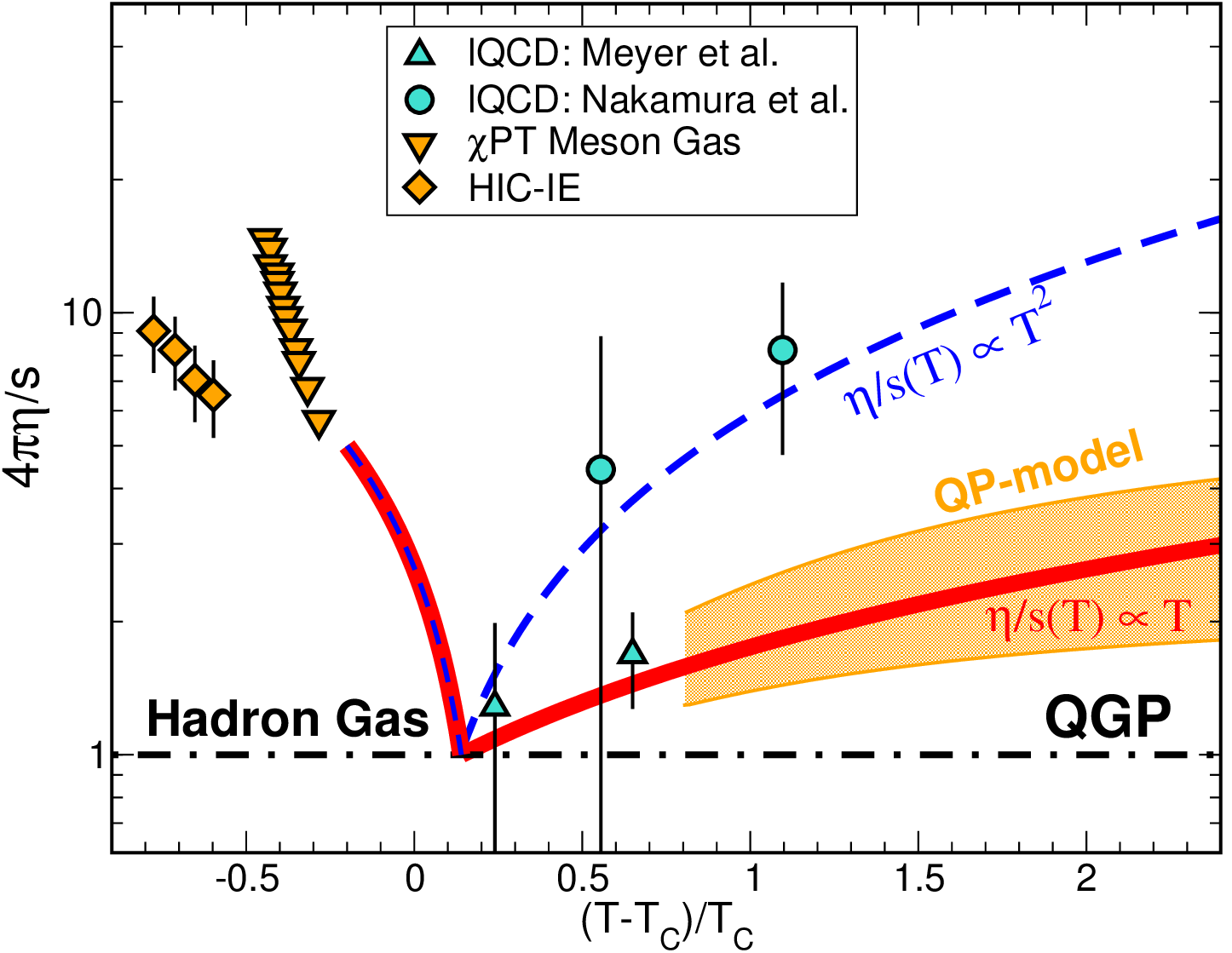}
\end{center}
\vspace{-0.6cm}
 \caption{Different temperature dependent parametrizations for $\eta/s$.
 The orange area takes into account the quasi-particle model predictions
 for $\eta/s$ \cite{Plumari:2011mk}.
 The different lines indicate different T dependencies assumed in
 the simulation of heavy-ion collision. Symbols are
 as in the legend. See the text for more details.}
 \label{Fig:etas_T}
\end{figure}

From both chiral perturbation theory for a meson gas \cite{Prakash:1993bt,Chen:2007xe} as well
as a transport analysis in uRQMD \cite{Demir:2008tr} and
the phenomenological analysis performed for heavy-ion collisions at intermediate
energy (HIC-IE) \cite{Schmidt:1993ak,Danielewicz:2009eu}
indicate a quite high value $4\pi \eta/s \geq 6$  for hadronic matter at a temperature $T < T_c =  \rm 165 \,MeV$,
see triangles and diamonds in Fig. \ref{Fig:etas_T}.
At higher T, first data on lattice QCD \cite{Meyer:2007ic} are compatible with $4\pi \eta/s \sim 1-2$ at
$T \sim T_c$ \cite{Meyer:2007ic}, see up-triangles in Fig. \ref{Fig:etas_T},
even if uncertainties are too large and the calculations have been performed only in the quenched approximation,
i.e. only in the limit of infinite quark masses.

It has been emphasized by both the STAR Collaboration at RHIC \cite{Adamczyk:2012ku} and the
ALICE at LHC  \cite{Aamodt:2010cz,Aamodt:2010pa} (in agreement with
the measurement done also by CMS \cite{Chatrchyan:2012ta} and ATLAS
\cite{ATLAS:2011ah}) that surprisingly the $v_2(p_T)$ appears to be
invariant in the very wide colliding energy range of $62.4 \rm \, GeV \leq \sqrt{s_{NN}} \leq 2.76 \rm \, TeV $.
Such an observation appears to be quite surprising at first sight because one would expect
that lowering the energy the contribution of hadronic matter over the evolution of the expanding
matter plays an increasing role in damping the $v_2(p_T)$, as indeed observed for the momentum averaged
$\langle v_2 \rangle$ \cite{STAR_PHENIX,Adcox:2004mh}.
It therefore appears a key question to answer the reason for the invariance of $v_2(p_T)$.

In this Letter, we show that the invariance of $v_2(p_T)$ in the energy range
$62.4 \rm \, GeV \leq \sqrt{s_{NN}} \leq 2.76 \rm \, TeV $ is caused by a fall and rise of the $\eta/s(T)$
as one would expect if the created matter undergoes a phase transition.
We employ a transport kinetic theory approach, developed to perform realistic simulations of HICs
keeping the local $\eta/s (T)$, to analyze the impact of different
temperature dependences assumed for the expanding matter.
Our study reveals that the invariance of $v_2(p_T)$ at varying
colliding energies means that the $\eta/s(T)$ has a typical "U"
shape with a decreasing behavior from the hadronic
matter and a not too steep rise with temperature in the QGP.

{\em Transport at fixed $\eta/s$ - }
We have developed in the recent years
a Relativistic Boltzmann Transport (RBT) approach that, instead of focusing on specific
microscopic calculations or modelings for the scattering matrix,
fixes the cross section in order to have the wanted $\eta/s$.  This is not the usual approach
to transport theory that is generally employed by starting from cross sections and
mean fields derived in microscopic models.
The motivation for our approach is inspired by the success of the hydrodynamical approach
that has shown the key role played by the $\eta/s$.
Therefore on one hand we use the RBT equation as an approach converging
to hydrodynamics for small scattering relaxation time $\tau \sim \sigma\rho$ (small $\eta/s$).
On the other hand the RBT equation is naturally valid also at large $\eta/s$ or $p_T>> T$
(explored in the present work) in contrast to hydrodynamics, and avoids uncertainties
in the determination of the viscous correction, $\delta f$, to the distribution function $f(x,p)$,
that usually becomes quite large at $p_T > \rm 1.5 \, GeV$ \cite{Dusling:2009df}.

To study the expansion dynamics with a certain $\eta/s(T)$, we determine locally in space and time
the total cross section $\sigma_{tot}$ according to the Chapmann-Enskog theory. For a
pQCD inspired cross section, $d\sigma/dt \sim \alpha_s^2/(t-m_D^2)^2$,
typically used in parton cascade approaches
\cite{Zhang:1999rs,moln02,Ferini:2008he,Greco:2008fs,Plumari_njl,Xu:2004mz,Xu:2008av},
this gives:
\begin{equation}
\eta/s =\frac{1}{15} \langle p\rangle \, \tau_{\eta}=
\frac{1}{15}\frac{ \langle p\rangle}{ g(a) \sigma_{tot} \rho} \,,
\label{eq:etas_CE}
\end{equation}
where $a=m_D/2T$, with $ m_D$ being the screening mass regulating the angular dependence
of the cross section $\sigma_{tot}$, while $g(a)$ is the proper function accounting for
the pertinent relaxation time $\tau_{\eta}^{-1}=g(a) \sigma_{tot} \rho$ associated
to the shear transport coefficient and given by:
\begin{eqnarray}
g(a)=\frac{1}{50}\! \int\!\! dyy^6
\left[ (y^2{+}\frac{1}{3})K_3(2 y){-}yK_2(2y)\right]\!
h\left(\frac{a^2}{y^2}\right),
\label{g_CE}
\end{eqnarray}
with $K_n$-s being the Bessel functions and
the function $h$ is relating the transport cross section to the total one
$\sigma_{tr}(s)= \sigma_{tot} \, h(m_{D}^2/s)$
and $h(\zeta)=4 \zeta ( 1 + \zeta ) \big[ (2 \zeta + 1) ln(1 + 1/\zeta) - 2 \big ]$.

The maximum value of $g$, namely $g(m_D \rightarrow \infty)=2/3$, is reached for isotropic cross section
and Eq.(\ref{eq:etas_CE}) reduces to the relaxation time approximation
with $\tau^{-1}_{\eta} =\tau^{-1}_{tr} = \sigma_{tr} \rho$.
We have shown in Ref. \cite{Plumari:2012ep} that Eq.(\ref{eq:etas_CE})
correctly describes the $\eta/s$ of the system
in the range of interest and it is in good agreement with the Green-Kubo formula.
We notice that in the regime where viscous hydrodynamics applies
the specific microscopic details of the cross section are irrelevant, and ours is the
only effective way to employ transport theory to simulate a fluid at a given $\eta/s$.

We solve the RBT equation with the constraint that $\eta/s(T)$ is fixed during
the dynamics of the collisions in a way similar to \cite{Molnar:2008jw},
but with an exact local implementation as described
in detail in \cite{Ferini:2008he}.
From Eq.(\ref{eq:etas_CE}) the cross section $\sigma_{tot}(\rho,T)$
determining the wanted value $\eta/s$ is given by:
\begin{equation}
\sigma_{tot}=\frac{1}{5}\frac{T}{\,g(T/m_D)\rho} \frac{1}{\eta/s}
\label{eq:eta_s2}
\end{equation}

\vspace{-0.2cm}

In our calculation the initial condition is longitudinal boost invariant flow, but the dynamical
evolution is 3D+1. For studying $v_2$ this approximation is adequate, although for other
collective flow phenomena, like rotation or turbulence \cite{hydro1,hydro2} more realistic
initial conditions would be necessary.
The initial $dN/d\eta$ have been chosen in order to reproduce the final $dN_{ch}/d\eta(b)$ at
mid rapidity as observed in the experiments at RHIC and LHC energies \cite{Alver:2010ck,Aamodt:2010cz}.
The partons are initially distributed according to the Glauber model in coordinate space.
In the momentum space the distribution is thermal up to $p_T=2 \rm \,GeV$ and
at larger $p_T$ we include the spectrum of non-quenched minijets according to standard
NLO-pQCD calculations.
In order to fix the maximum temperature in the center of the fireball, $T_{m0}$, we assume that it scales with
the collision energy according to the relation
\begin{equation}
\frac{1}{\tau A_T}\frac{dN_{ch}}{d\eta} \propto T^{3}\,,
\end{equation}
and for the initial time, $\tau_0$, we ensure that it satisfies the uncertainty
relation between the initial average thermal energy and the initial time by $T_{m0}\tau_0 \approx 1$.
Combining these two relations one has
\begin{equation}
\frac{T(\sqrt{s_1})}{T(\sqrt{s_2})}=\sqrt{\frac{dN_{ch}/d\eta(\sqrt{s_1})}{dN_{ch}/d\eta(\sqrt{s_2})}}
\end{equation}
as commonly done in hydrodynamical studies \cite{Kestin:2008bh}.
Thus at 62.4 GeV, 200 GeV and 2.76 TeV the maximum initial
temperature, $T_{m0},$ has the values 290 MeV, 340 MeV and 560 MeV respectively.
Once the maximum temperature is fixed, the local temperature profile scales with the energy
density $T(\vec{r})=T_{m0}\big( \epsilon(\vec{r})/\epsilon(0) \big)^{1/4}$

%
%

%

\begin{figure}[h]
\begin{center}
\includegraphics[width=8.3 cm]{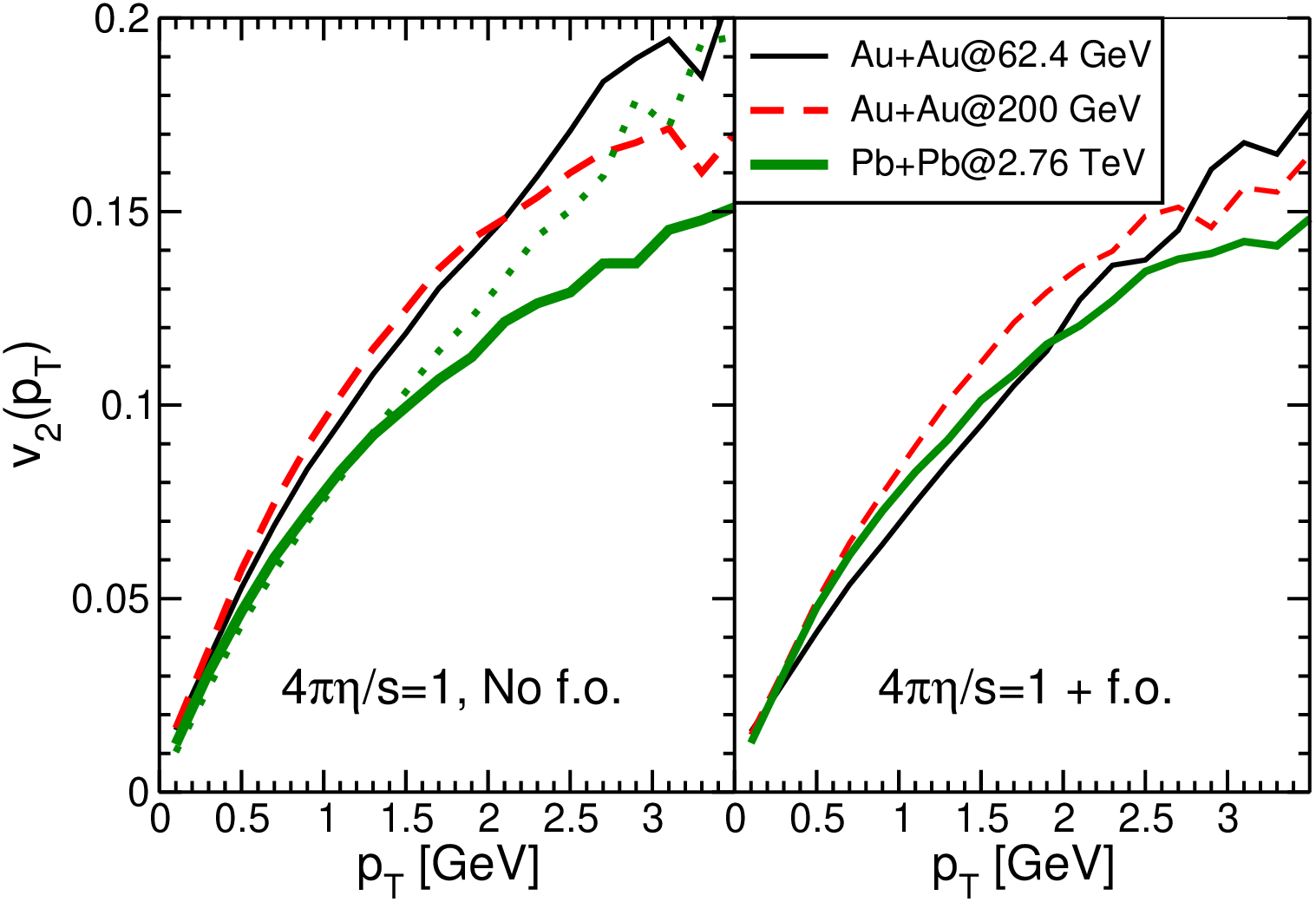}
\end{center}
\vspace{-0.6cm}
 \caption{Differential elliptic flow $v_2(p_T)$ at mid rapidity for $10\%-20\%$ collision centrality.
The thin solid line, the dashed line  and the thick solid line refer to:
$Au+Au$ at $\sqrt{s}= 62.4$ GeV and
$\sqrt{s}= 200$ GeV and $Pb+Pb$ at $\sqrt{s}= 2.76$ TeV, respectively.
On the left panel, the results of the simulation with $4\pi \eta/s = 1$ during
the whole evolution of the system, while on the right panel
with the inclusion of the kinetic freeze-out are shown. See text for more details. The dotted line is the result excluding
minijets at $\sqrt{s}= 2.76$ TeV.}
\label{Fig:v2_1}
\end{figure}
\vspace{-0.2cm}

In the following we will discuss the impact of the temperature dependence of $\eta/s$,
In the cross-over region that we identify as the region just below the knee in the $\epsilon/T^4$ curve \cite{Borsanyi:2010cj},
i.e. at $ \epsilon < \rm 1.5 \,GeV/fm^3$ and/or $T<T_0=1.2T_c$,
the $\eta/s$ should increase linearly at decreasing T matching the estimates from chiral perturbation theory
for a high temperature meson gas \cite{Prakash:1993bt,Chen:2007xe}, shown by triangles in Fig. \ref{Fig:etas_T}.
We notice that the last are also comparable to the estimate of $\eta/s$ extrapolated from heavy-ion collisions
at intermediate energies (HIC-IE diamonds in Fig. \ref{Fig:etas_T}), even if one should consider that they refer
to a matter with higher baryon chemical potential $\mu_B$.
As discussed above, due to the large error bars in the lQCD results for $\eta/s$ it is not possible to infer a clear
temperature dependence in the QGP phase. We have considered two cases. One with a linear
dependence $4\pi \eta/s=T/T_0$ (red solid line in Fig. \ref{Fig:etas_T})
in agreement with the indication of lQCD calculation in  Ref. \cite{Meyer:2007ic}
(up-triangles) as well as with quasi-particle model prediction (orange band) suggesting $\eta/s \sim T^{\alpha}$ with
$\alpha \approx 1 - 1.5$ \cite{Plumari:2011mk,Bluhm:2010qf}. The other one with a quadratic
dependence $4\pi \eta/s=3.64(T/T_0-1)+(T/T_0)^2$ (blue dashed line) resembling the lQCD in quenched approximation
in Ref. \cite{Nakamura:2004sy} as given also in \cite{Niemi:2012ry}.
We also consider a common case of a constant $\eta/s$ at its conjectured minimum value $1/4\pi$.

In Fig. \ref{Fig:v2_1}, we plot the results for the $v_2(p_T)$ for the three different beam energies at RHIC and LHC
at the same centrality $10-20\%$ and for two different $\eta/s(T)$ as described in the following.
On the left panel, the results are shown for $\eta/s=1/4\pi$
all over the evolution of the system. We see clearly that such a case would not predict an invariance of
$v_2(p_T)$ up to the LHC energy, but would generate a breaking up of about $20\%$.
It is interesting also to notice that remaining in the regime of RHIC
one would have indeed an approximate scaling of $v_2(p_T)$ that however results to be misleading
on a wider energy scale.
This makes us to understand the importance of a wide beam energy scan (BES) up to LHC energy.

In Fig. \ref{Fig:v2_1} (left panel), we show by the dotted line
the effect of modifying the initial $p_T$ distribution at LHC
discarding the minijets. We clearly see that just this change
can significantly affect the $v_2(p_T)$ at least at $p_T>1.5 \rm\, GeV$.
Therefore, behind the observed scaling there is an implicit role
of the initial $p_T$ distribution that has to be correctly implemented including the non-equilibrium at increasing
$p_T$. Generally, the effect is that a stiffer distribution (like the one of mini-jets) produces a smaller $v_2(p_T)$
even if the lifetime of the fireball and the assumed $\eta/s$ are unchanged.

In Fig. \ref{Fig:v2_1} (right panel), we show the pattern of $v_2(p_T)$ when an increase of $\eta/s(T)$ is assumed
in the cross-over region, solid or dashed line in Fig. \ref{Fig:etas_T}. We label this case as $4\pi\eta/s=1+ f.o.$ to emphasize that accounting for
the increase of $\eta/s$ in the cross-over region naturally realize a freeze-out because it implies a smooth switching-off
of the scattering cross section.
For such a case, we see that the $v_2(p_T)$ at different energies becomes more similar, even if still far
from a scaling as observed experimentally.  The main reason behind a more similar $v_2(p_T)$ is that
the fireball created at 62.4 GeV is more affected by the increase of $\eta/s$ in the cross-over region,
while the system created at LHC is practically not affected at all.
This of course makes the $v_2(p_T)$ more similar between RHIC and LHC, even if at 200 GeV
it still remains larger because it is less affected by the increase of $\eta/s(T)$ at low T compared
to the 62.4 GeV case.

The different impact of $\eta/s(T)$ on $v_2(p_T)$  is determined by the different initial temperature
and consequent lifetime of the stage at $T>T_c$. In fact at RHIC energies such a lifetime is about $4 - 6 \rm\, fm/c$
while at LHC about $10 \rm\,fm/c$, in agreement with HBT results \cite{ALICE:HBT}.
Therefore at RHIC the elliptic flow has not enough time to fully
develop in the QGP phase, while at LHC the lifetime is long enough to let the $v_2$ develop almost completely
in the QGP phase and the increase of $\eta/s(T)$ at low T becomes irrelevant, as also found in Ref. \cite{Niemi:2011ix}.

From this reasoning one has the hint that an invariant $v_2(p_T)$ can be caused by the specific T-dependence of
$\eta/s$ that balances the suppression due to
the viscosity above and below $T_c$ where a minimum in $\eta/s$ should occur.
We have considered the T-dependence  of $\eta/s$ also at $T>T_0$  in the QGP stage. In one case, we consider an
$\eta/s(T)$ rapidly (quadratically) increasing in the QGP phase, see the dashed blue line in Fig. \ref{Fig:etas_T} and in the other case
a linear increase with T in agreement with lQCD data of Ref. \cite{Meyer:2007ic}, corresponding to the red solid line in Fig. \ref{Fig:etas_T}.

\begin{figure}[ht]
\begin{center}
\includegraphics[width=8.3 cm]{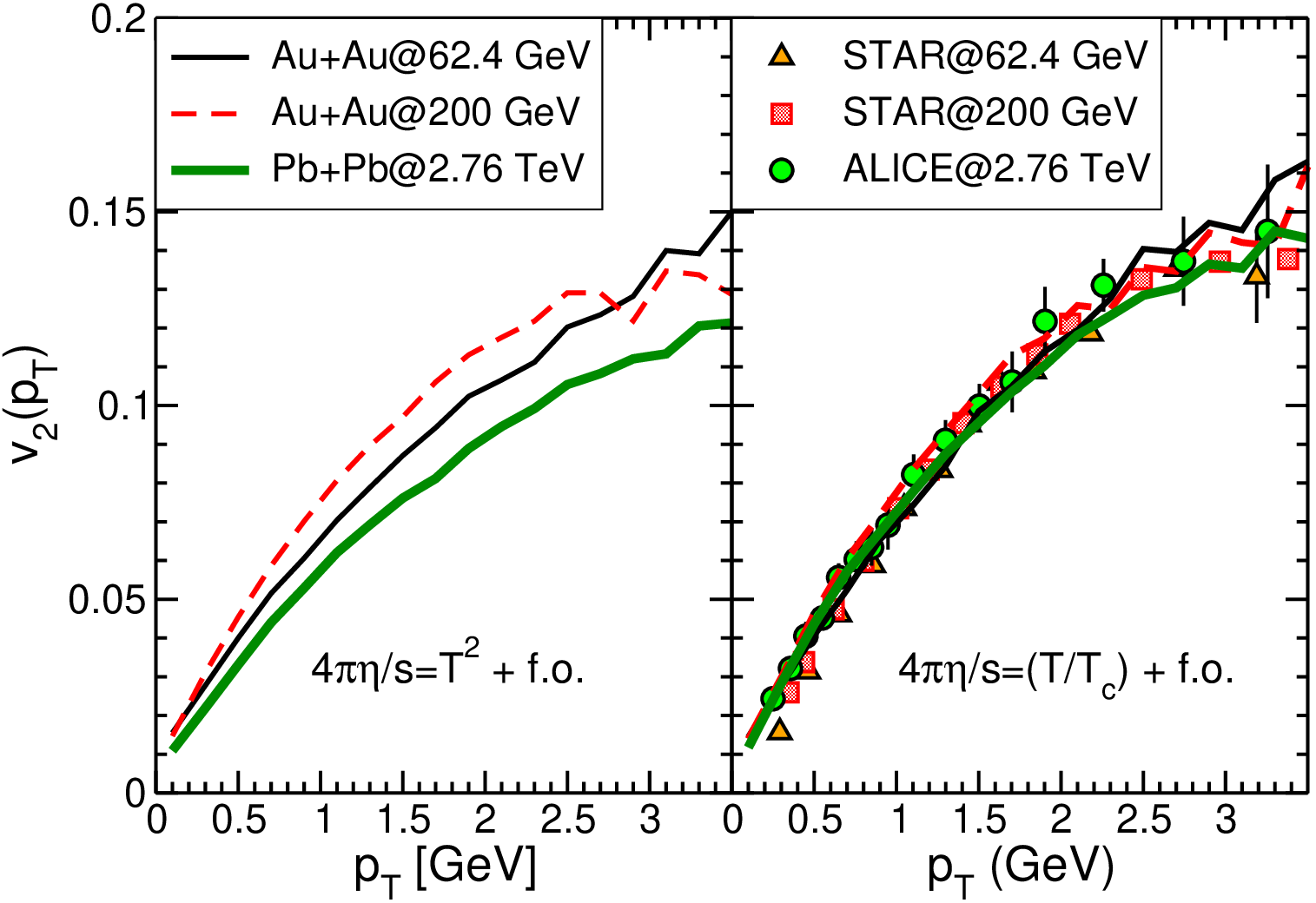}
\end{center}
\vspace{-0.4cm}
\caption{As in Fig. \ref{Fig:v2_1} but for two different $\eta/s(T)$: a quadratic dependence on the left panel and a linear one
on the right panel. In the right panel data for $v_2[4]$ measured by STAR and ALICE collaborations
\cite{Aamodt:2010pa,Adamczyk:2012ku} are shown by different symbols as indicated in the legend.}
\label{Fig:v2_2}
\end{figure}

As we can see comparing the right panel of Figs. \ref{Fig:v2_1} and  the left one of Fig. \ref{Fig:v2_2}
the rapidly increasing $\eta/s(T)$
affects more the system created at LHC and this generates again a larger splitting of the
$v_2(p_T)$ among the different energies. This again means that also a strong T-dependence in the QGP phase
is in contrast with the observed $v_2(p_T)$ invariance, essentially it would make the system at LHC too much
viscous.
Finally we consider $4\pi\eta/s=T/T_0$, according to the solid red line in Fig. \ref{Fig:etas_T}.
The results in the right panel of Fig. \ref{Fig:v2_2} are also
compared with the experimental results for the $v_2[4]$ measured at RHIC
and LHC energy, data taken by \cite{Aamodt:2010pa,Adamczyk:2012ku}.
We can see that in such case there is an almost perfect invariance of $v_2(p_T)$ (within a $5\%$) in agreement
with what is observed in the experimental data shown by the different symbols in Fig. \ref{Fig:v2_2} (right panel).
The main effect is that with a mild increase of $\eta/s(T)$ at $T>T_0$ the elliptic flow at LHC energies goes up
reaching the higher $v_2(p_T)$ obtained at lower energies.

From a comparison with
the first case considered in Fig. \ref{Fig:v2_1} (left panel), we understand that to have an invariant $v_2(p_T)$
it is essential also the rise of the $\eta/s(T)$ in the hadronic or cross-over region that significantly
reduces the $v_2$ at low RHIC energy.
Therefore only a fall and rise of the $\eta/s(T)$ can account for a $v_2(p_T)$
almost invariant going 62.4 GeV to 2.76 TeV.
Furthermore, a comparison of the results for all the $\eta/s(T)$ considered
for RHIC at 200 AGeV shows that this is the case less affected by the T dependence of $\eta/s$.
Paradoxically this is the case more thoroughly studied by means of both hydrodynamical and transport approaches
till now.  However, we also notice that the impact of the T dependence of $\eta/s$
on the $v_2(p_T)$ is anyway quite weak and
we have been able to educe the necessity of non vanishing T dependence only thanks to
the experimental observation of the $v_2$ scaling and exploiting
a direct comparison in a quite wide range of colliding energy.
Still one should notice that we have been able to discriminate a constant $\eta/s$ or a strong dependence like the
quadratic one that for the maximum initial temperature at LHC would mean about a factor ten larger $\eta/s$ respect
to the conjectured $1/4\pi$ lower bound.
The main reason is probably that even if the $\eta/s$ is large at larger temperatures this is really
relevant only for the inner side of the fireball
at the beginning of the expansion while most the elliptic flow is anyway formed later and more in the
peripheral region of the fireball where the temperatures or energy densities
are quite similar both as a function centralities or beam energy. However, we have shown that still
a comparative analysis at different beam energies is able to reveal an important information on the $\eta/s(T)$,
namely the necessity of a "U" shape of $\eta/s(T)$ with a minimum slightly
above $T_C$, as expected when the matter undergoes a cross-over \cite{Csernai:2006zz,Lacey:2007na}.
Therefore this finding provides a nice evidence for the phase transition of matter created in uRHIC's.

\vspace{-0.1cm}

Our result shows also that a BES with relativistic heavy-ion collisions allows to infer
key properties of the created matter that are not otherwise accessible and hence BES is far from being
a mere repetition of a similar experiment.
We notice that in the present study we find that the increase of $\eta/s(T)$
below the QGP phase is consistent with previous studies
at intermediate energies which further supports the reliability of the study of nuclear matter through HIC's.
In this respect an expert reader could wonder why we studied the elliptic flow from BES starting
from 62.4 GeV while data are available from 7.7 GeV. We mention that this is due to the fact that
at lower energy the baryon chemical potential $\mu_B$ is no longer negligible, as pointed
out also in \cite{Lacey:2007na}. This would probably imply that a non-vanishing
vector potential is acting as discussed in Ref. \cite{Song:2012cd}.

Our study is a seminal analysis of the information that we can obtain
from the huge experimental
efforts of the last decade, and one would expect that on a similar footing more stringent constraints can be
obtained from an analysis of higher harmonics like $v_3$. As this implies the development
of initial state fluctuations in the transport approach and this is out of the reach of the present
study, even if a work in such a direction is already under development.

We find that for the different beam energies considered the suppression of the elliptic flow due to the viscosity
of the medium has different damping coming from the hadronic or QGP phase depending on the average energy of the system.
In particular, we observe that at LHC the elliptic flow is much less damped by the hadronic phase allowing a better study
of the QGP properties.
Moreover we have found that going from RHIC to LHC energies it is possible to have a nearly invariant $v_2(p_T)$
only if the $\eta/s$ has a fall and rise with a minimum around the transition $T_c \sim 165 \rm\, MeV$ a behavior
expected when there is a phase transition or a cross-over.

{\em Acknowledgments -}
This work is partially supported by the ERC-StG under the Grant QGPDyn No. 259684.
L.P. Cs. thanks for enlightening discussions and kind hospitality at the
Helmholtz International Center for FAIR.


\end{document}